\documentclass[pra,twocolumn,showpacs]{revtex4-1}
\usepackage{graphicx,epstopdf}
\usepackage{amsmath}
\usepackage{amssymb}
\usepackage{epsfig}
\usepackage{epstopdf}
\usepackage{subfigure}
\usepackage{enumerate}

\usepackage[colorlinks=true, citecolor=blue, urlcolor=blue ]{hyperref}

\begin{document}
\title{Effect of long-range interactions on multipartite entanglement in Heisenberg chains}
\author{Sudipto Singha Roy\(^{1,2}\) and Himadri Shekhar Dhar\(^{3,4}\)}
\affiliation{\(^1\)Instituto de F{\'i}sica T{\'e}orica  UAM/CSIC, C/ Nicol{\'a}s Cabrera 13-15, Cantoblanco, 28049 Madrid, Spain\\
\(^2\)Department of Applied Mathematics, Hanyang University (ERICA), 55 Hanyangdaehak-ro, Ansan, Gyeonggi-do, 426-791, Korea\\
\(^3\)Institute for Theoretical Physics, Vienna University of Technology, Wiedner Hauptstra{\ss}e 8-10/136, 1040 Vienna, Austria\\
\(^4\)Physics Department, Blackett Laboratory, Imperial College London, SW7 2AZ London, UK
}
%

\begin{abstract}
It is well known that the notions of spatial locality are often lost in quantum systems with long-range interactions, as exhibited by emergence of phases with exotic long-range order and faster propagation of quantum correlations. 
We demonstrate here that such induced ``quasi-nonlocal" effects do not necessarily translate to growth of global entanglement in the quantum system. 
By investigating the ground and quenched states of the variable-range, spin-1/2 Heisenberg Hamiltonian, we observe that the genuine multiparty entanglement in the system can either increase or counterintuitively diminish with growing range of interactions.
The behavior is reflective of the underlying phase structure of the quantum system and provides key insights for generation of multipartite entanglement in experimental atomic, molecular and optical physics where such variable-range interactions have been implemented.
\end{abstract}

\maketitle
\section{Introduction}
In recent years, there has been considerable interest in investigating the physical properties related to quantum systems with long-range interactions \cite{Lahaye2009, Peter2012, Gong2016, Zeiher2017}. 
This is primarily in response to the significant developments made in experimental atomic, molecular and optical (AMO) physics \cite{Bloch2008,Saffman2010,Douglas2015}, where such interactions can be implemented in a well controlled setting \cite{Schauss2012,Britton2012,Yan2013,Islam2013}. 
These studies have led to a flurry of exciting new physical phenomena 
\cite{Hauke2013,Sciolla2011,Zunkovic2018, Richerme2014, Jurcevic2014, Schachenmayer2013, Buyskikh2016, Bruno2001, Laflorencie2005, Lobos2013, Maghrebi2017,  Smith2016, Jaschke2017, Neyenhuis2017, Eldredge2017}, for instance, propagation of correlations faster than the Lieb-Robinson bound \cite{Hauke2013,Richerme2014,Jurcevic2014}, 
emergence of exotic long-range order \cite{Laflorencie2005,Lobos2013,Bruno2001,Maghrebi2017}, and
dynamical phase transitions \cite{Sciolla2011,Zunkovic2018}.
{Most of these phenomena arise in the presence of long-range interactions due to the breakdown of ``quasi-locality"~\cite{Eisert2013,Gong2017} 
(cf.~\cite{Luitz2019}). In this context, quasi-locality affirms the existence of a non-relativistic spatial light-cone within which most of the causal information travels with the finite Lieb-Robinson velocity~\cite{Lieb1972}. Any correlations or response to local fluctuations appear to be strongly suppressed at small distances away from this light-cone boundary \cite{Hastings2006,Nachtergaele2006}, which may not be the case when long-range interactions are present.}
Importantly, the loss of quasi-locality can lead to nontrivial distribution of quantum entanglement  \cite{Horodecki2009}, 
which over the years has been established as an important resource in implementation of various quantum information and computation protocols \cite{Ekert1991,Bennett1993,Bennett1992} (also see \cite{Nielsen2000}). While recent studies have { focused} on the growth of entanglement between two parties in a variable-range interacting system \cite{Schachenmayer2013, Buyskikh2016}, the effect of emergent quasi-nonlocality due to long-range interactions on the global entanglement of these systems remains elusive \cite{Pappalardi2018}.
Here, we address this void by investigating the multipartite entanglement in the ground and quenched states of quantum many-body systems with long-range interactions.

It is known that entanglement jointly distributed among many parties has richer features \cite{Vidal2000,Verstraete2002}, which has allowed for the design of sophisticated protocols such as cryptographic conference \cite{Horne1992, Bose1998} and multiparty quantum communication \cite{Karlsson1998,Bandyopadhyay2000,Rigolin2005,Yeo2006,Agrawal2006,Ghose2016}.
Multiparty entangled states 
are also intrinsic resources in implementation of novel quantum computation models such as measurement-based quantum computation \cite{Raussendorf2001}. In the past decade, notable progress in experimental physics has allowed for the efficient creation and manipulation of multiparty entanglement \cite{Eibl2004,Prevedel2009,Gao2010,Pan2012,Yao2012}. This opens up the exciting potential for harnessing systems with tunable range of interactions for physical realization of these quantum protocols. Moreover, multiparty entanglement is also an important characteristic quantity in the study of critical phenomena in many-body systems \cite{Wei2005,Cui2008,Orus2008,Giampaolo2013,Stasinska2014,Hofmann2014,Roy2017,Pezze2017}.

In this work, we consider a spin-1/2 Heisenberg chain with spin interactions that follow a power-law decay (${1}/{r^{\alpha}}$). Efficient implementation of such variable interactions has been possible with recent developments in AMO physics, in particular with cold atoms \cite{Douglas2015} where the parameter $\alpha$ can be tuned. Other systems include Rydberg atoms \cite{Schauss2012}, trapped ions \cite{Britton2012} and polar molecules \cite{Yan2013}. 
Incidentally, it is known that such power-law decay in Heisenberg chain can lead to breakdown of quasi-locality \cite{Maghrebi2017,Luitz2019}. An important ramification of this is that the area law no longer bounds the entanglement entropy \cite{Gong2017} (cf.~\cite{Hastings2007,Eisert2010}), especially for $\alpha \leq 1$. In the same vein, intuitively, one would expect that the spatial nonlocal effects induced by the long-range interactions will result in quantum phases with enhanced global entanglement.
To explore this further, 
we characterize the multiparty entanglement in both the ground and quenched states of the considered Hamiltonian.
We observe a clear dichotomy between two different regimes, depending on whether the interactions in the $x$--$y$ spin plane are antiferromagnetic (AFM) or ferromagnetic (FM). 
{We note that while the ground states} in the FM regime have enhanced multiparty entanglement for increased range of interactions in the system, counterintuitively, for the AFM regime the global entanglement weakly diminishes. 
{Interestingly, we note that this is no longer the case when quantum states are quenched with such long-range interactions. Here, we start from a completely separable or product spin state and switch on the interactions. The subsequent growth of multipartite entanglement in the time-evolved system is then numerically analyzed.}
%
%
{Here, we observe that the} AFM interactions with long-range are more favorable towards the growth of multiparty entanglement, in contrast to the FM interactions, where the growth appears almost independent of the range of interactions. Thus, our findings clearly demonstrate that long-range interaction selectively enhances quantum resources, such as global entanglement in the system, and this is important for experimental efforts to generate entanglement and implement {quantum information and computation} protocols using systems with variable-range interactions.

The paper is arranged as follows. We introduce the spin-1/2 Heisenberg chain with long-range interactions in Sec.~\ref{model}. Our measure of genuine multipartite entanglement and its computaion is discussed in Sec.~\ref{sec:GGM_def}. In Sec.~\ref{sec:results}, we analyze the genuine multiparty entanglement in the ground states of the long-range model. The growth of entanglement under quantum quench is then investigated in Sec.~\ref{sec:quench}, before we end with a final discussion on the results in Sec.~\ref{sec:discussions}.

\section{\label{model} Model}

\begin{figure}[t]
\includegraphics[width=3.4in,angle=00]{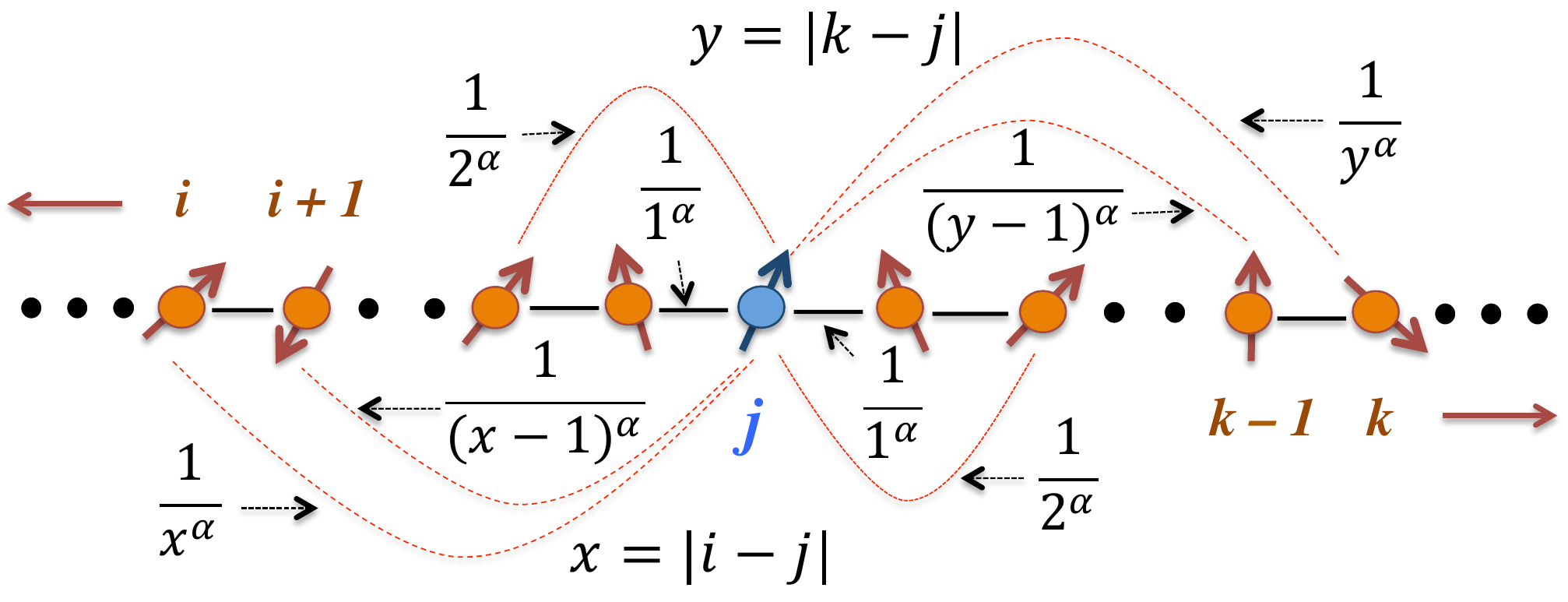}
\caption{Schematic of a quantum spin chain with interactions that follow a power-law decay, ${1}/{r^{\alpha}}$. The black (bold) and red (dotted) lines show the short- and long-range coupling between the $j^{th}$  and the other spins in the quantum system. 
}
\label{fig1}
\end{figure}

We start by introducing the physical system of our interest, the one dimensional (1D) quantum spin lattice, consisting of spin-1/2  particles, coupled via long-range interactions with power-law decay. The Heisenberg Hamiltonian governing such a system can be written as
\begin{eqnarray}
\mathcal{H}=\sum_{i<j}\frac{1}{|i-j|^{\alpha}}(J_x\sigma^x_i \sigma^x_{j}+J_y\sigma^y_i\sigma^y_{j}+\Delta\sigma^z_i \sigma^z_{j}),
\label{Long_Ham}
\end{eqnarray}
where $\alpha\geq0$ is the continuous exponent that controls the long-range interaction. $J_x$ and $J_y$ are the coupling constants along the $x$ and $y$ spin axes, respectively, 
and $\Delta$ is the anisotropy along the $z$-direction. Here, $\sigma^m$'s are  the Pauli spin matrices ($m\in \{x,y,z\}$). For $J_x, J_y < 0$, the interaction in the $x$--$y$ plane is ferromagnetic, while for $J_x, J_y > 0$, we obtain the antiferromagnetic coupling. The above Hamiltonian, in the presence of long-range interactions, has a rich phase diagram. For $J_x = J_y = -1$, an exotic continuous symmetry breaking (CSB) phase emerges \cite{Maghrebi2017} for low values of $\alpha$, apart from the known \emph{XY} and AFM phases observed in the short-range Hamiltonian. This is a true hallmark of the quasi-nonlocal effect and {the change in effective dimensionality of the system induced by long-range interactions. This is due to the fact that spontaneous breaking of continuous symmetry typically appears only in higher dimensional spin lattices and is otherwise forbidden  in low-dimensional systems by the Mermin-Wagner theorem~\cite{Mermin1966} (also see Ref.~\cite{Maghrebi2017}).} 
For $\alpha=\infty$, the model reduces to the short-range Heisenberg model with nearest-neighbor (NN) interactions. A schematic of a 1D long-range quantum spin system of arbitrary size and with power-law decay of interactions is provided in Fig.~\ref{fig1}. In our study, 
we consider periodic boundary conditions for the spin chain. { In order to do that we always choose the interaction corresponding to the shortest distance, $|i-j|$, from one site and another in the periodic spin chain.} We note that at $\Delta=0$, for $J_x = J_y$, the above Hamiltonian gives us the long-range \emph{XX} model, which we analyze in our study. Moreover, for $\Delta = J_x = J_y$ and $\alpha=2$, the model reduces to the exactly solvable Haldane-Shastry model \cite{Gaudiano2008,Shastry1988,Haldane1988}.
%


\section{\label{sec:GGM_def} Measure of genuine multiparty entanglement}

Before going into the detailed analysis of the entanglement properties of the ground and quenched states of the long-ranged Heisenberg model, we begin by defining the genuine multiparty entanglement of a quantum state. We note that  there exists several equivalent definitions and measures of multiparty entanglement in the literature \cite{Horodecki2009}. In our work, we are mainly { focused} on the  genuine multiparty entanglement of a quantum system \cite{Chiara2018}, which is defined as follows: {\it An $N$-party  pure quantum state, $|\psi\rangle_N$, is said to be genuinely multiparty entangled if it cannot be written as a product in any bipartition}. In other words, a genuine multiparty entangled state is entangled across all  bipartitions of the system \cite{Goldbart2003, Blasone2008, Sen2010}. In order to estimate this quantity in $|\psi\rangle_N$, we consider the generalized geometric measure (GGM) \cite{Sen2010}, which is a computable measure of genuine multiparty entanglement of a state. It is defined as an optimized  distance of the given quantum state, $|\psi\rangle_N$, from the set of all states that are not genuinely multiparty entangled. This can be mathematically expressed as
$
\mathcal{G}(|\psi\rangle_N)=1-\Lambda_{\max}^2(|\psi\rangle_N),
$
where $\Lambda_{\max} (|\psi\rangle_N ) = \max | \langle \chi|\psi\rangle_N |$, with the maximization being over all  such pure quantum state $|\chi\rangle$ that are  not genuinely multiparty entangled. Following some simplifications, one can derive an equivalent expression for the above equation, given by \cite{Sen2010}
\begin{equation}
\mathcal{G} (|\psi \rangle_N ) =  1 - \max \{\lambda^2_{ A : B} |  A \cup  B = \{s_1,\ldots, s_N\},  A \cap  B = \emptyset\},
\label{GGM}
\end{equation}
where \(\lambda_{A:B}\) is  the maximal Schmidt coefficient  of $|\psi\rangle_N$, in the  bipartition 
$A:B$. The measure is then optimized over all possible  bipartitions of the state, $|\psi\rangle_N$, and for spin-1/2 or qubit systems, takes values in the range: $ 0 \leq \mathcal{G} (|\psi\rangle_N ) \leq 1/2$. In recent years, GGM has been used to characterize genuine multiparty entanglement in strongly-correlated systems, including quantum spin liquids \cite{Dhar2013,Roy2016}, doped spin lattices \cite{Roy2017b,Das2018}, and other many-body systems \cite{Prabhu2011, Jindal2014, Mishra2016, Biswas2014, Sadhukhan2017}.

We note that the computation of GGM in many-body quantum systems requires access to the complete state of the system and all its reduced density matrices.  In general, for a quantum system with $N$ number of sites, the number of such reduced density matrices is given by $\sum_{i=1}^{N/2} {N \choose i} $, which increases exponentially with the size of the system. Moreover, in the presence of long-range interactions, there are no known analytical or approximate methods, such as tensor-networks or matrix product states, which can be used to compute GGM, as is the case in several short-range models (see Refs.~\cite{Dhar2013,Roy2016,Roy2017b,Roy2018}).
Therefore, in our case, we are restricted to exact numerical solutions for small, finite spin chains. In our work, we have considered systems with up to $N$ = 20 spins, and use diagonalization and propagation methods based on the Krylov subspace and Lanczos algorithm. {To mitigate the effect of unstable finite-size effects in the presence of long-range interactions, we have also checked the qualitative consistency of our main results against smaller system-sizes.}


\section{\label{sec:results} Multiparty entanglement in the ground state}

We now study the variation of genuine multiparty entanglement ($\mathcal{G}$) in the {ground states} of the spin-1/2 Heisenberg chain, with long-range interactions, given by Eq.~(\ref{Long_Ham}). Towards that aim, we consider two distinct regimes emanating from the Hamiltonian, i) The ferromagnetic regime, with interactions in the $x$--$y$ plane given by $J_x = J_y = -1$, and ii) the antiferromagnetic, with $J_x=J_y=1$. Here, we consider only the antiferromagnetic interactions along the $z$-axis, i.e.,$0 \leq \Delta/J \leq 2$ ($J$ = $|J_x| = |J_y|$), where there always exists a distinct gap between the lowest energy values. The long-range interaction in the system is controlled through the exponent $\alpha$, which is varied in the integer range, $1 \leq \alpha \leq 10$. We exclude the extreme points corresponding to systems with infinite interactions ($\alpha = 0$) or strictly NN interactions ($\alpha = \infty$). 
%
%

\begin{figure}[t]
\includegraphics[width=3in,angle=00]{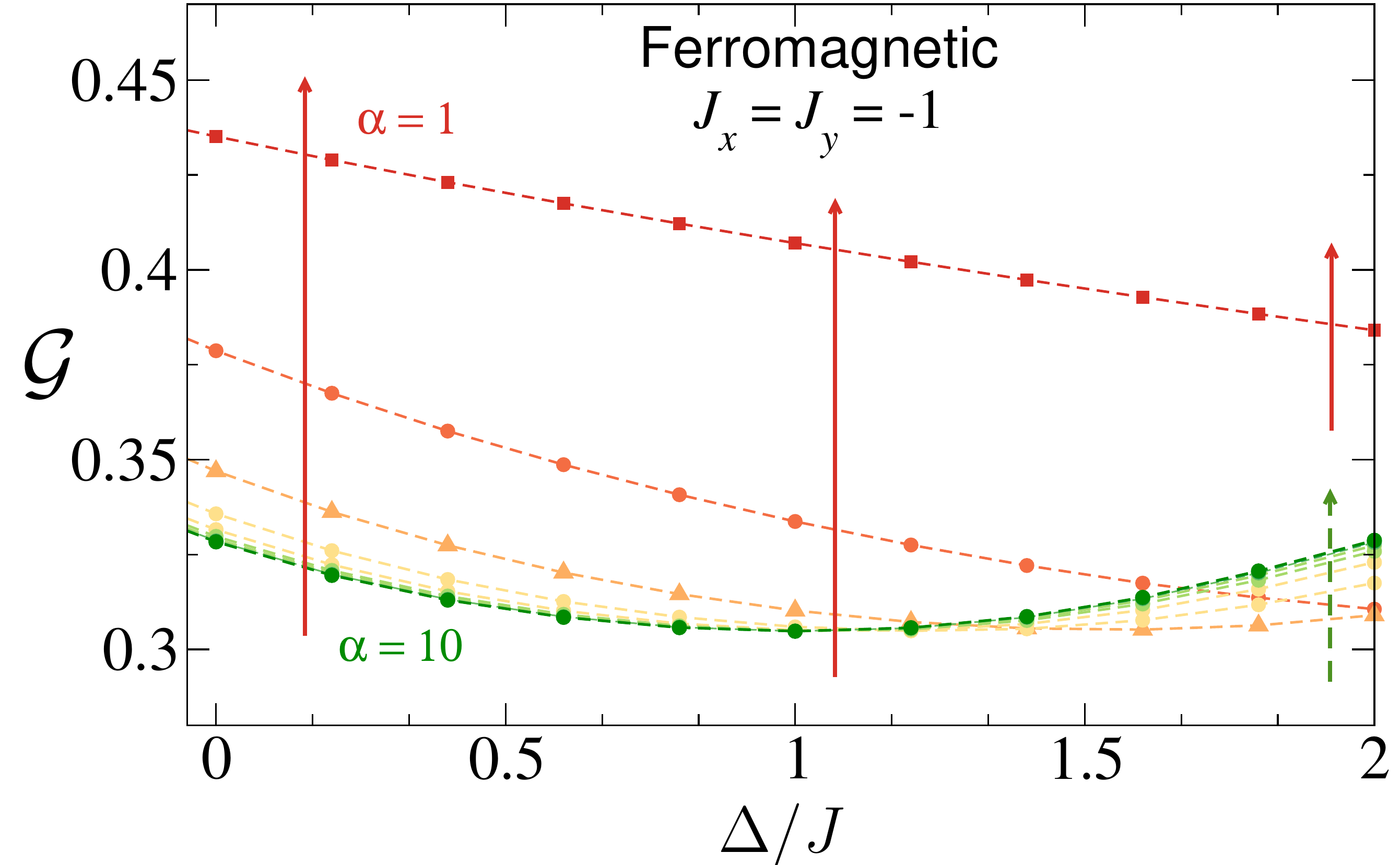}
\caption{(Color online.) Variation of genuine multipartite entanglement. Here, we consider a Heisenberg chain with $N = 20$ spins, and FM interactions ($J_x=J_y=-1$) in the 
$x$--$y$ plane. 
The plot shows the variation in $\mathcal{G}$ with the parameter $\Delta/J$, where $J$ = $|J_x| = |J_y|$, for ten different integer values of the exponent $\alpha$, ranging from $\alpha = 1$ (red-squares) to $\alpha = 10$ (green-circles). Here, the dashed lines are fits to the plotted data points. The regime $\Delta=0$ corresponds to the \emph{XX} model which also mimics the result obtained for Heisenberg chain at low $\Delta$. 
We note that the plots for higher $\alpha$ values are very close together, which shows that the short-range character is reached fairly quickly.
Moreover, the solid-red vertical arrows highlight specific parameter regimes where $\mathcal{G}$ increases with decreasing $\alpha$ (or increasing long-range interactions), whereas the dashed-green vertical arrows show regions where $\mathcal{G}$ increases but now for increasing $\alpha$ (or decreasing long-range interactions).
Both the axes are dimensionless.
}
\label{fig2}
\end{figure}

We start with the FM regime, and consider the case where the interaction is defined by $\alpha = 10$, with variable anisotropy between the $x$--$y$ and $z$ directions. 
We note that the system is already short-range for $\alpha = 10$. 
In Fig.~\ref{fig2}, we note that the ground state is genuinely multiparty entangled for all values of the anisotropy parameter, $\Delta/J$ ($0\leq\Delta/J\leq 2$), with the minimum $\mathcal{G}$ at the point, $\Delta/J = 1$. As the range of interaction is increased, by decreasing $\alpha$, the genuine multipartite entanglement increases monotonically, for $\Delta/J < 1$. 
Subsequently, at higher values of anisotropy ($\Delta/J > 1$) the local minimum shifts to larger values of $\Delta/J$ for decreasing values of $\alpha$. Moreover, there is a cross-over  
between $\mathcal{G}$ values of different $\alpha$, which gives rise to an interesting regime where the shortest-range ($\alpha = 10$) and the longest-range of interactions ($\alpha = 1$) generate ground states with higher global entanglement than the intermediate range of interactions. 
%
%
More importantly, $\mathcal{G}$ remains the highest at $\alpha = 1$, and gradually decreases with increasing $\Delta$. Therefore, as long-range interactions in the system {increase} there is an expected hike in the genuine multipartite entanglement {in the ground state} of the ferromagnetic spin-1/2 Heisenberg Hamiltonian. 

\begin{figure}[t]
\includegraphics[width=3in,angle=00]{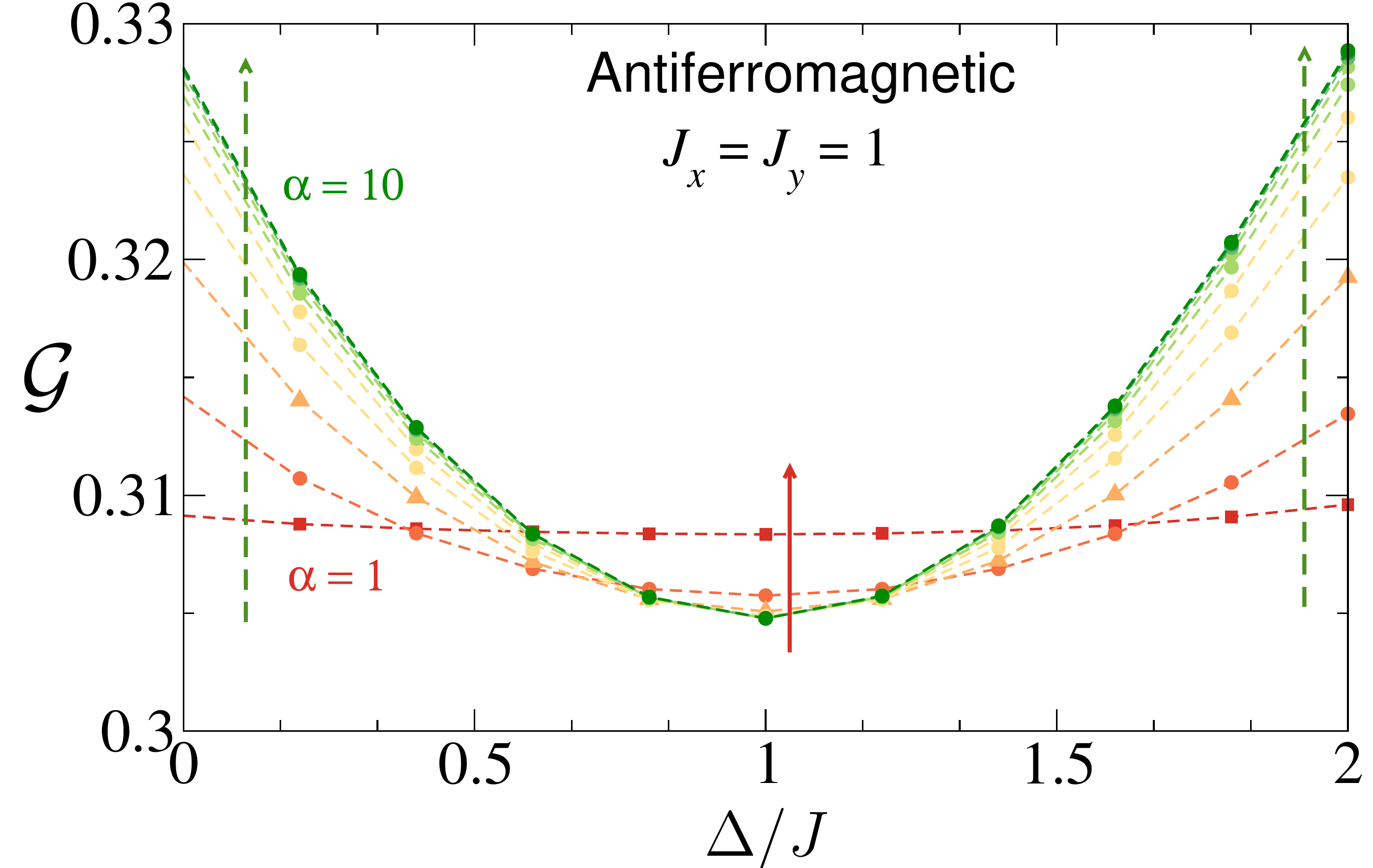}
\caption{(Color online.) Variation of genuine multipartite entanglement. Here, we consider a Heisenberg chain with $N = 20$ spins, and AFM interactions ($J_x=J_y=1$) in the $x$--$y$ plane. 
The plot shows the variation in $\mathcal{G}$ with the  parameter $\Delta/J$ (where, $J$ = $|J_x| = |J_y|$), for ten different integer values of the exponent $\alpha$, ranging from $\alpha = 1$ (red-squares) to $\alpha = 10$ (green-circles). 
Once more, the dashed lines are just fits to the plotted data points and both the axes here are dimensionless.
The red and green vertical arrows here imply the same behavior as outlined in Fig.~\ref{fig2}.
}
\label{fig3}
\end{figure}

In the AFM regime, the situation is drastically different. For the short-range interaction ($\alpha = 10$), the genuine multiparty entanglement of the ground state is minimum at $\Delta/J = 1$, with a distinct symmetry around the point, as shown in Fig.~\ref{fig3}. 
In contrast to the FM regime, $\Delta/J = 1$ is the local minima of $\mathcal{G}$, for all values of $\alpha$, {although in the vicinity of this point the multipartite entanglement increases for more long-range interactions, which is similar to the FM regime.}
However, away from this point, $\mathcal{G}$ decreases as the long-range interaction in the system is increased. This intriguing behavior of genuine multipartite entanglement is in direct contrast to the behavior of the system in the FM regime, and implies a negative interdependence between global entanglement and long-range induced nonlocal effects in the system.
This is significant from the perspective of physical implementation of quantum protocols where multiparty entanglement is an important resource. In the AFM regime, long-range interactions appear to be detrimental to generating large entangled states, as compared to the FM regime.

{The difference in the behavior of the genuine multipartite entanglement between the FM and AFM regimes, can be partly explained using a heurtistic description of the ground state in these regimes based on our numerical simulations. Here, the competition between different ground state configurations in interacting many-body systems gives rise to the phenomena of entanglement-frustration \cite{Dawson2019} (also see Refs.~\cite{Roy2017,Jindal2014}), which can potentially define the complex behavior of entanglement in our model. In the long-range interaction model that we consider, the ground state can be written as a superposition between two stable, but competing configurations, such that $|\psi_g\rangle = a~|\psi_\mathrm{N}\rangle + b~|\phi_\mathrm{\bar{N}}\rangle$.
Here, $|\psi_\mathrm{N}\rangle$ is the state arising due to the N{\'e}el order at $\Delta > 1$ (for $J$ = 1). It is expected that at large $\Delta$, the ground state will be closer to the N{\'e}el state for both the AFM and FM model. For large $\alpha$, this is simply given by $|\psi\rangle$ = $|\uparrow\downarrow\uparrow\cdots\downarrow\rangle$, where $\{\uparrow, \downarrow\}$ are the eigenstates of  $\sigma_z$. However, the complementary configuration $|\psi'\rangle$ = $|\downarrow\uparrow\downarrow\cdots\uparrow\rangle$ is also a likely ground state at large $\Delta$. Hence, at dominant $\Delta$ values, frustration ensures that, $|\psi_\mathrm{N}\rangle = \beta_1|\psi\rangle + \beta_2|\psi'\rangle$.  The parity symmetry of $\mathcal{H}$ results in $\beta_1 = \pm \beta_2 = 1/\sqrt{2}$, which ensures $|\psi_\mathrm{N}\rangle$ is maximally multiparty entangled.
On the other hand, $|\phi_\mathrm{\bar{N}}\rangle$ refers primarily to the non-N{\'e}el configurations in the ground state, which are orthogonal to both $|\psi\rangle$ and $|\psi'\rangle$. 
These states are significant in regimes where $\Delta$ is not large and seem to arise from the \emph{XY} terms in the Hamiltonian.  
However, unlike $|\psi_\mathrm{N}\rangle$, the entanglement properties of  $|\phi_\mathrm{N}\rangle$ are a priori not known.

While the above description is intuitively appealing for regimes that correspond to either large or small values of the anisotropy parameter, numerical analysis suggests that it can also provide a broad picture of the ground state for intermediate values of $\Delta$.
By investigating the quantum fidelity of the ground state to $|\psi\rangle$ and $|\psi'\rangle$, one can deduce that the overall entanglement of the ground state is  dependent on the trade-off between the states $|\psi_\mathrm{N}\rangle$ and $|\phi_\mathrm{\bar{N}}\rangle$, i.e., the ratio $a/b$.
In the AFM case, two distinct regimes emerge for all $\alpha$, symmetric around the point $\Delta = 1$, viz. the region with $a >b$ (for $\Delta > 1$) and the one with $b > a$ (for $\Delta < 1$). We call these the N{\'e}el and non-N{\'e}el regimes. Later, we discuss how these regimes closely correspond to the AFM and \textit{XY} phases respectively. In the N{\'e}el regime, we observe that the ratio $a/b$ not only increases with $\Delta$ but also with $\alpha$, resulting in higher entanglement for shorter-range interactions.  
Interestingly, in the non-N{\'e}el regime the opposite behavior is observed. Here, it can be numerically shown that the ratio $b/a$ increases for decreasing $\Delta$ values, 
resulting in more entanglement close to $\Delta = 0$. However, $b/a$ also increases with increasing $\alpha$, which again leads to higher entanglement in short-range systems. 
This allows a distinct symmetry in multiparty entanglement to emerge around the vicinity of $\Delta = 1$ (in the region, $0\leq\Delta\leq2$), for all values of $\alpha$ in the AFM model, but with short-range interactions leading to more entanglement, as shown in Fig.~\ref{fig3}. Things look more interesting in the FM case, where the effects of long-range interactions become more prominent. Firstly, for $\alpha > 1$, the transition from the N{\'e}el to the non-N{\'e}el regime is no longer centered at $\Delta = 1$, apart from the short-range cases ($\alpha > 6$). The different points of transition on $\Delta$ increases for decreasing $\alpha$. This allows for cross-over between the multiparty entanglement corresponding to different values of $\alpha$. Secondly, and more importantly, there is no N{\'e}el to non-N{\'e}el transition for $\alpha = 1$ in the FM case, at least within the considered parameter regime. Therefore, the ground state always corresponds to high values of $b/a$ (in the non-N{\'e}el regime) and has high multiparty entanglement compared to other values of $\alpha$ (see Fig.~\ref{fig2}).} 

{Incidentally, we note that in the short-range limit (i.e. $\alpha =10$), due to the $SU(2)$ invariance of the Hamiltonian, the AFM and FM model turns out to be the same at $\Delta=\pm 1$. Moreover, in the short-range limit, the FM and AFM models here are also connected via local unitary operations (spin-flip operations at alternate sites in the spin chain) that keep the global entanglement unchanged. This is reflected in Figs.~\ref{fig2} and \ref{fig3}, where the plots for multiparty entanglement ($\mathcal{G}$) at $\alpha = 10$ are almost same for both the FM and AFM cases.}

\begin{figure}[t]
\includegraphics[width=3.5in,angle=00]{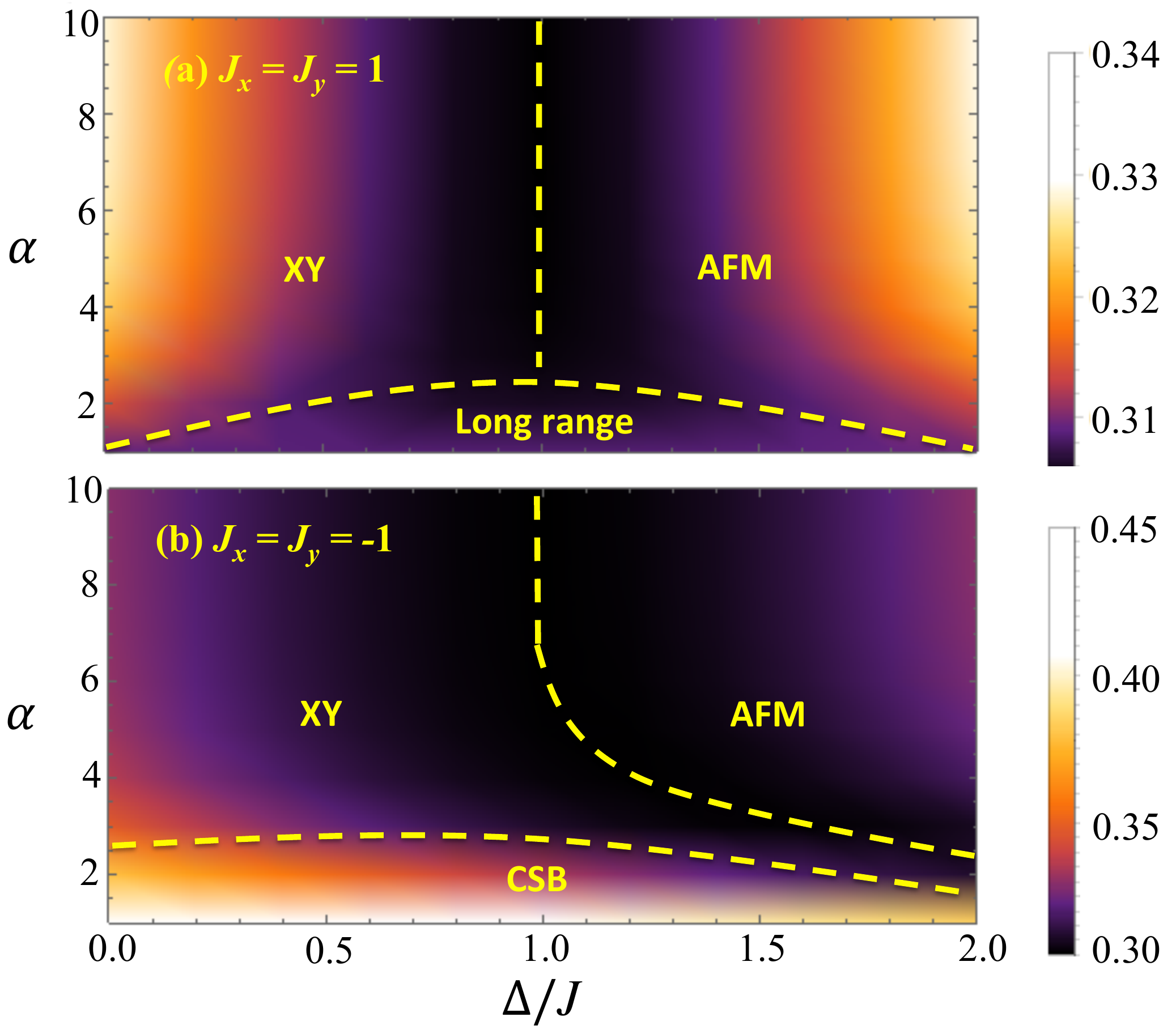}
\caption{(Color online.) Variation of  genuine multiparty entanglement ($\mathcal{G}$) in the ground states  of the long-range Heisenberg chain consisting of $N = 20$ spins  in the $\Delta/J-\alpha$ plane for (a) AFM and (b) FM interactions in the $x$--$y$ plane. Here, $J$ = $|J_x| = |J_y|$.
We note here that the dashed yellow lines represent the extremal (minima) points of $\mathcal{G}$. However, they only serve as a visual aid to deconstruct the known quantum phases of the model (see Ref.~\cite{Maghrebi2017}). Both the axes and the color bar in the figures are dimensionless.
}
\label{fig4}
\end{figure}

%

{The dichotomy} in the behavior of genuine multipartite entanglement in the ground state of FM and AFM regimes of the Heisenberg Hamiltonian is closely related to their respective phase structures. In Fig.~\ref{fig4}, we show that the genuine multipartite entanglement is able to deconstruct the different phases in these regimes, as has been established in earlier work \cite{Maghrebi2017}. For large $\alpha$, the phases are similar to their counterparts corresponding to NN spin-1/2 Heisenberg chain, with two distinct phases: the \emph{XY} spin liquid phase and the AFM Ising-like phase. Figures~\ref{fig4}(a)-\ref{fig4}(b), shows how $\mathcal{G}$ distinctly highlights these phases in both the AFM and FM regimes, respectively. 
We note that the ferromagnetic phase corresponding to $\Delta < -1$ is not shown in the diagram, as $\mathcal{G}$ cannot be uniquely computed for degenerate ground states. The anomalous behavior arises as $\alpha$ is decreased and one enters the quasi-nonlocal regime. For the AFM case, a regime of relatively weak entangled  phase appears, with lower values of $\mathcal{G}$. 
In contrast, in the FM regime, the continuous symmetry breaking phase emerges with decreasing $\alpha$ ($\alpha \lesssim 2$) \cite{Maghrebi2017}, which is marked by a region of high genuine multiparty entanglement. Therefore, in terms of the phase diagram, the increase in genuine multipartite entanglement with increasing long-range interactions is related to the \emph{XY}--CSB phase transition in the FM regime. In addition to this, at the truly long-range interaction limit ($\alpha\sim1$), the ground state mostly remains in the CSB phase, which apparently does not decrease  quickly even when the anisotropy and N{\'e}el order increase with  $\Delta$. 


\section{\label{sec:quench}Generating entanglement through quantum quench}

We now look at how genuine multipartite entanglement can be generated through a  quantum quench mediated {by the variable-range interactions in either the FM or AFM regimes of the spin Hamiltonian}. In particular, we start with a product or completely separable initial state of the system, given by $|\psi\rangle_{in} = \prod_{i}^N|\phi\rangle_i$, where $|\phi\rangle_i = \frac{1}{\sqrt{2}}(|0\rangle_i+|1\rangle_i)$. Here, $|0\rangle_i$ and $|1\rangle_i$ are the  eigen states of $\sigma_i^z$.  
The initial states here can be thought to be ground states of some local Hamiltonian acting identically on all the spins. For the quantum quench, the long-range interactions are instantaneously switched on in the spin system. Subsequently, the initially separable quantum state rapidly evolves in time leading to potential growth of multiparty entanglement in the system. 
We note that the quench performed in our study is motivated from the perspective of various quantum information and computation protocols, where entanglement is necessary for successful implementation of the protocol. To this end, in our quench process we begin with a completely separable product state, which is a resource-less state, and wish to generate useful resource (entanglement) in the system.
Our main aim here is to see whether the presence of long-range interactions in the Heisenberg Hamiltonian can generate higher entanglement or quantum resource in these quenched states {as compared to process that only invoke short-range interactions during the quench. 

\begin{figure}[h]
\includegraphics[width=3.5in,angle=00]{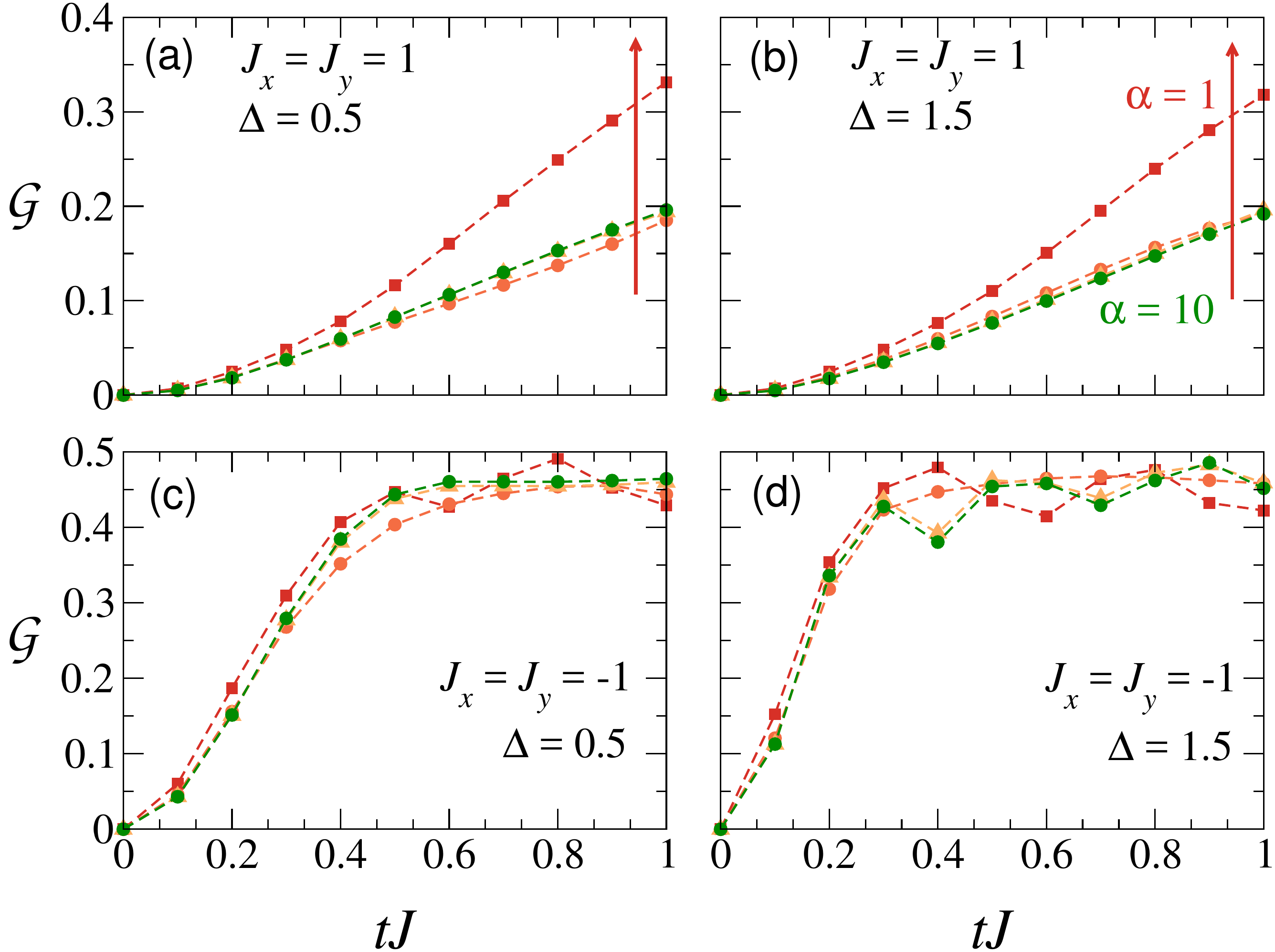}
\caption{(Color online.) Genuine multiparty entanglement after a quantum quench. The growth of $\mathcal{G}$ in a system consisting of $N = 12$ spins after a quantum quench of the initially product state, $|\psi\rangle_{in}$, for (a)-(b) AFM and (c)-(d) FM interactions in the $x$--$y$ plane. Here, $J$ = $|J_x| = |J_y|$ and the plots correspond to $\alpha =$ 1 (red-squares), 2 (green-circles), 5 (yellow-triangles), 10 (orange-circles). The red vertical arrow here implies the same behavior as outlined in Fig.~\ref{fig2}.
The axes in the above figures are all dimensionless.
}
\label{fig5}
\end{figure}

The initial state is subjected to a  quantum quench and coherently evolves to $|\psi(t)\rangle = \exp(-i\mathcal{H}t)|\psi\rangle_{in}$. Subsequently, we measure how much GGM is generated in the quenched state, i.e., we calculate $\mathcal{G}(|\psi(t)\rangle)$. 
We are interested in the parameter regimes away from $\Delta = 1$, where the dichotomy between {the ground states in the  FM and AFM cases} appears to be the most distinct. Figure~\ref{fig5}, shows the evolution of the state after the quench. Surprisingly, for the quenched dynamics, long-range interactions ($\alpha = 1$) appears to play a strong role in the growth of multipartite entanglement when $|\psi\rangle_{in}$ is quenched in the AFM regime. In contrast, the generation of multipartite entanglement in the FM regime is almost independent of the range of interactions in the system. This implies that highly entangled quantum states can be generated through quenching in the FM regime even in the absence of any significant long-range interactions.
Therefore, in quenched dynamics long-range interactions seem to affect the multiparty entanglement favorably in the AFM regime, while remaining ambivalent in the FM regime. This is converse to the outcome that was observed in the ground state phases of the system.  

\section{\label{sec:discussions}Discussion}

In this work, we have demonstrated how the quasi-nonlocal effect induced by long-range  interactions in many-body systems, selectively affects the multipartite entanglement of the system. By investigating different ground state phases of the spin-1/2 Heisenberg Hamiltonian we observed that multiparty entanglement can be enhanced or counterintuitively, can reduce as the range of interactions are increased. In particular, these opposing effects were observed for two distinct ground states phases depending on whether the interaction in the $x$--$y$ plane was ferromagnetic or antiferromagnetic. While the global entanglement is expectedly boosted with more quasi-nonlocal effects for the FM regime, in contrast, long-range interactions {appear to} act detrimentally in the AFM case. {A possible reason for the unexpected behavior in the AFM regime, as we observed in our ground state analysis, is the entanglement-frustration arising from N{\'e}el and non-N{\'e}el terms, which appears to favor more global entanglement in the short-range limit. In the FM case, no transition occur between the N{\'e}el and non-N{\'e}el regimes for long-range interactions, leading to higher entanglement.}


Interestingly, the observed dichotomy in these regimes was intriguingly different while considering the generation of multiparty entanglement through quenched dynamics of initially separable states. Here, long-range interactions allow for robust growth of global entanglement in the AFM regime, in contrast to the FM regime, where there is no perceptible advantage in using longer interactions in the quenched dynamics. Overall, our results clearly demonstrate that the system in the ferromagnetic interaction regime is more susceptible to allow significant global entanglement for both short- and long-range interactions.


Our findings provide significant insights for physical implementation of quantum protocols where multiparty entanglement is the necessary resource, such as measurement based computation or secure multiparty communication.
With recent technological breakthroughs in experimental atomic, molecular and optical physics, where the systems often contain tunable long-range interactions, it is essential to determine the optimal range of interactions that will allow for maximal global entanglement in these system, which can then be harnessed in the quantum protocol.

\acknowledgments
The authors thank R. Fazio for useful discussions at ICTP, Trieste. HSD acknowledges funding by the Austrian Science Fund (FWF), project no. M 2022-N27, under the Lise Meitner programme of the FWF.

\end{document}